\documentclass{article}
\usepackage{lnfprep}
\usepackage{graphicx}
\usepackage{subfigure}
\usepackage{amsmath}
\usepackage{amssymb}

\def\beq{\begin{equation}}   \def\eeq{\end{equation}}
\def\bea{\begin{eqnarray}}   \def\eea{\end{eqnarray}}

\newcommand{\gsim}{\lower.7ex\hbox{$ \;\stackrel{\textstyle>}{\sim}\;$}}
\newcommand{\lsim}{\lower.7ex\hbox{$ \;\stackrel{\textstyle<}{\sim}\;$}}

\def\c2{CLEO~II.V}

\def\d0d0{ D^0\bar{D}^0 }
\def\p0p0{ P^0\bar{P}^0 }

\def\qp2{ \Bigl| \frac{q}{p} \Bigr|^2 }
\def\pq2{ \Bigl| \frac{p}{q} \Bigr|^2 }

\def\ps2s{  \psi(2S) }
\def\q2{ $q^2$ }
\def\cm2s1{ $\,{\rm cm}^{-2} {\rm s}^{-1}$}
\def\d0{D_2^{*0}}
\def\d+{D_2^{*+}}

\newcommand{\Header}{
  \begin{tabular}{rl}
  \hspace{-.4cm}
      &
    \renewcommand{\arraystretch}{0.5}
    \renewcommand{\arraystretch}{1}
  \end{tabular}
  \vskip 1cm
  \begin{flushright}
  \renewcommand{\arraystretch}{0.5}
    \begin{tabular}{r}
      {\underline{LNF-05/31 (P)}}\\    
      {\small 22 dicembre 2005} \\      
    \end{tabular}
  \end{flushright}
  \renewcommand{\arraystretch}{1}
  \vskip 1 cm
  }

\begin{document}
\begin{titlepage}
\title{
  \Header
  {\LARGE  \textsc{\textmd {TWO- AND THREE-DIMENSIONAL RECONSTRUCTION AND
ANALYSIS OF THE STRAW TUBES TOMOGRAPHY IN THE BTEV EXPERIMENT}}
  }
}
\author{E.Basile(*), F.Bellucci (***), L.
Benussi, M. Bertani, S. Bianco, M.A. Caponero (**),  \\
D. Colonna (*), F. Di Falco (*), F.L. Fabbri, F. Felli (*), M.
Giardoni, A. La Monaca, \\
F.Massa (*), G. Mensitieri (***), B. Ortenzi, M. Pallotta, A.
Paolozzi (*), L.
Passamonti, \\
D.Pierluigi, C. Pucci (*), A. Russo, G. Saviano (*)\dag.\\
{\em Laboratori Nazionali di Frascati dell'INFN, v.E.Fermi 40 00044
Frascati (Rome) Italy } \\
\\
\\
F.Casali, M.Bettuzzi, D.Bianconi
\\
{\em University of Bologna and INFN, Bologna, Italy }
\\
~ \\
presented by F.~Massa at ICATPP05, Villa Olmo (Como) Italy 2005
}
\maketitle
\baselineskip=1pt

\begin{abstract}
\indent \indent A check of the eccentricity of the aluminised kapton
straw tubes used in the BTeV experiment is accomplished using X-ray
tomography of the section of tubes modules. 2 and 3-dimensional
images of the single tubes and of the modules are reconstructed and
analysed. Preliminary results show that a precision better than 40
$\mu$m can be reached on the measurement of the straw radii.
\end{abstract}

\vspace*{\stretch{2}}

\vskip 1cm
\begin{flushleft}
\begin{tabular}{l l}
  \hline\\
  $ ^{*\,\,\,\,\,\,}$ & \footnotesize{Permanent address: ``La Sapienza" University - Rome.} \\
  $ ^{**\,\,\,}$& \footnotesize{Permanent address: ENEA Frascati.} \\
  $ ^{***}$ & \footnotesize{Permanent address: ``Federico II" University - Naples.} \\
\end{tabular}
\end{flushleft}
\dag This work was supported by the Italian Istituto Nazionale di
Fisica Nucleare and Ministero dell'Istruzione, dell'Universit\`a e
della Ricerca. This work was partially funded by contract EU
RII3-CT-2004-506078.
\end{titlepage}
\pagestyle{plain}
\setcounter{page}2
\baselineskip=17pt

\section{  \textsc{Introduction}}
The BTeV experiment [1] uses straw tubes glued in modules and embedded in
a structure mechanically untensioned [2], where straw and microstrip detector
are integrated, allowing a minimum amount of materials. A check of the eccentricity
of the straw tubes and their position is accomplished using X-ray tomography of
the module sections.\\
\newline
\section{  \textsc{BTeV Detector}}
Experiment straw section is scanned orthogonal to the vertical axis
of the tomograph with 27 $\mu$m resolution. Data are initially
reduced in a numerical 8-bit matrix of 1024x1024 points, then
converted to an IMAQ Image of LabVIEW.\\
Figure 1 show the raw tomographic image of a straw tube section
(section$\sharp10$, 1024x1024 pixels, 27 $\mu$m/pixel, 256 values of
greys). Not all the tubes of the module are contained in the field
of view. Figure 1 also shows evident traces of glue, deposited on
the external surfaces of the straw tubes, especially in the points
of contact of adjacent tubes. Only the internal surfaces are well
defined in the images and, in order to detect and to measure
possible mechanical deformations of the tubes, this forces to study
the geometry of the these surfaces. Figure 2 shows the distribution
of the greys values of the previous figure.\\ An improvement of the
signal-to-noise con be obtained just setting an upper threshold of
the intensity, as it is shown in Fig.3, reporting the effect of a
threshold of 210 of the image of Fig.1. As an example, an arbitrary
straight of the pixels along this line is shown if Fig.4, where the
two peaks point out the positions of the crossing points (Edges) of
the line with the internal surfaces of the two adjacent tubes. The
transformation of the Edges coordinates from the line reference to
the image reference is easily obtained from the coordinates of the
line end points in the image reference. On this base, an automatic
procedure is defined in order to obtain the Edges of 14 tubes of a
section of the module.\\
The procedure is as follows. First an image containing a Region of
Interest is built, then the patterns recognising such a region are
extracted from the tomography. At the Edge pointed out on the base
of contrast figures, three orthogonal coordinates X,Y,Z, defined in
the tomograph system (Tomo reference), are attributed, where the Z
is common to all the Edges of the same section. This allows the
reconstruction of the 2 and 3-dimensional images of single tubes and
of the entire module.\\
We do not expect a perfect positioning of the module on the
tomograph reference plane, and even in the case of perfect
positioning we would not expect a perfect parallelism between the
tubes of the module. Therefore, the section Edges of each straw tube
are fitted to an ellipse. The centres of the ellispes of all
sections are in turn fitted to a straight line: the axis of the
straw tube.\\
Projecting the Edges of a section on the plane orthogonal to the
axis of its straw tube the contribution to an elliptical
configuration due to a not perfect vertically of the tube is
eliminated.
 \newline
\section{  \textsc{Results}}
We fit the data points to an ellipse, and define as Ellipse
Parameter the quantity
\begin{equation*}
    (\underline{PF_{1}}+\underline{PF_{2}})/2
\end{equation*}
where P is a point on the ellipse, $F_{1,2}$ are the ellipse foci,
and $\underline{PF_i}$ their distances.\\
In order to evaluated the amount of the mechanical deformation of
the straw tubes cross section from the expected circular shape, we
then fit the data to a circle. Figure 5 shows the standard deviation
of the histograms of the Edge radius, and the width of gaussian fit
to the Ellipse Parameter of the projected edges for the 14 straw
tubes analyzed. In the worst case the mechanical deformation
respect to the circular cross section have a distribution with a
standard deviation of about 1.5 pixel, corresponding to about
40$\mu$m, largely contained in the $100\mu$m specification in order
not to spoil the electric field inside the straw tube. The precision
of our technique in determining the variation of the straw
cross-section from circularity cam be estimated by the difference in
quadrature of the two variances in Fig.7, which is about 1.2 pixels
at most, corresponding to about $30\mu$m. The three-dimensional rendering of
slices reconstructed is shown in Figure 8.
 \newline
\begin{figure}[!htbp]
\begin{center}
  \includegraphics[width=5cm]{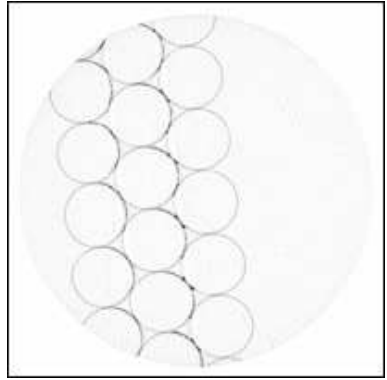}\\
  \caption{Example of raw tomography image of straw tubes module (1024x1024 pixel, $27\mu$/pixel)}
  \end{center}
\end{figure}

\

\begin{figure}[!htbp]
\begin{center}
  \includegraphics[width=5cm]{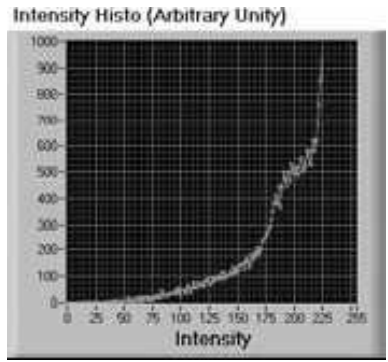}\\
  \caption{Grey intensity histogram of Fig.1}
  \end{center}
\end{figure}

\begin{figure}
\begin{center}
  \includegraphics[width=5cm]{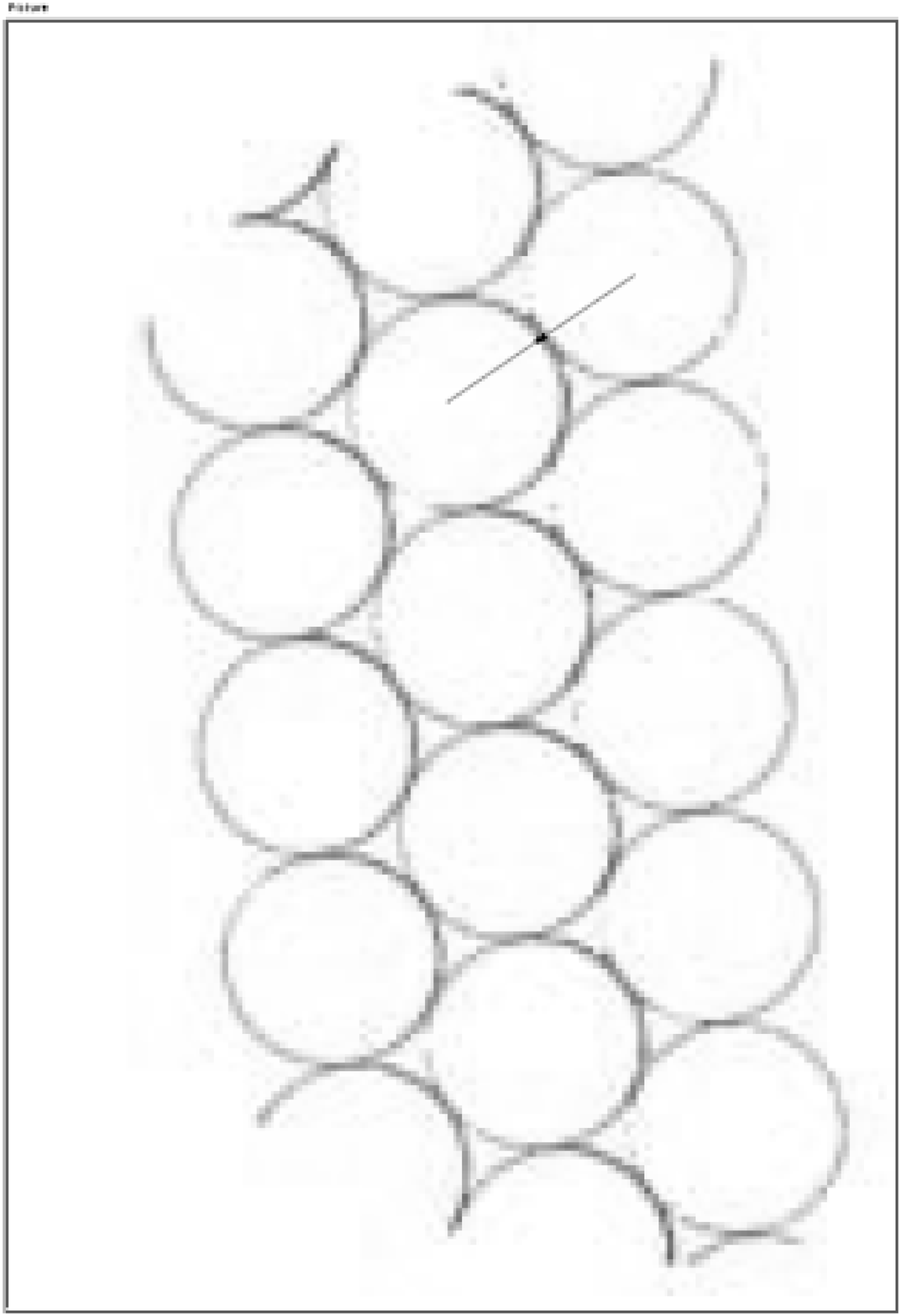}\\
  \caption{Same tomography of Fig.1 with a grey intensity threshold of 210. The intensity distribution
  along the segment is shown in Fig.4.}
  \end{center}
\end{figure}

\begin{figure}
\begin{center}
  \includegraphics[width=5cm]{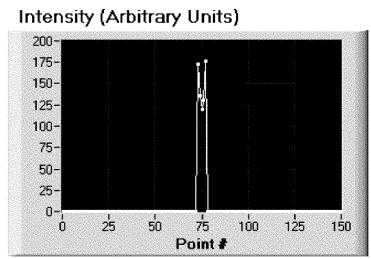}\\
  \caption{Intensity along the line of Fig.3.}
  \end{center}
\end{figure}

\begin{figure}
\begin{center}
  \includegraphics[width=5cm]{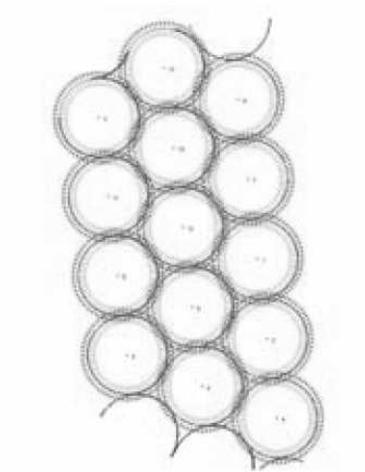}\\
  \caption{Edges of 14 straw tubes of the section$\sharp10$ of the module. Each
  Edge is chosen as the first met in the arrow direction, from the inside to the outside of the
  Region of Interest, pointing out to the internal surface of the straw tube. Circles fitting
  the Edges, their centers and order number are also shown}
  \end{center}
\end{figure}

\begin{figure}
\begin{center}
  \includegraphics[width=5cm]{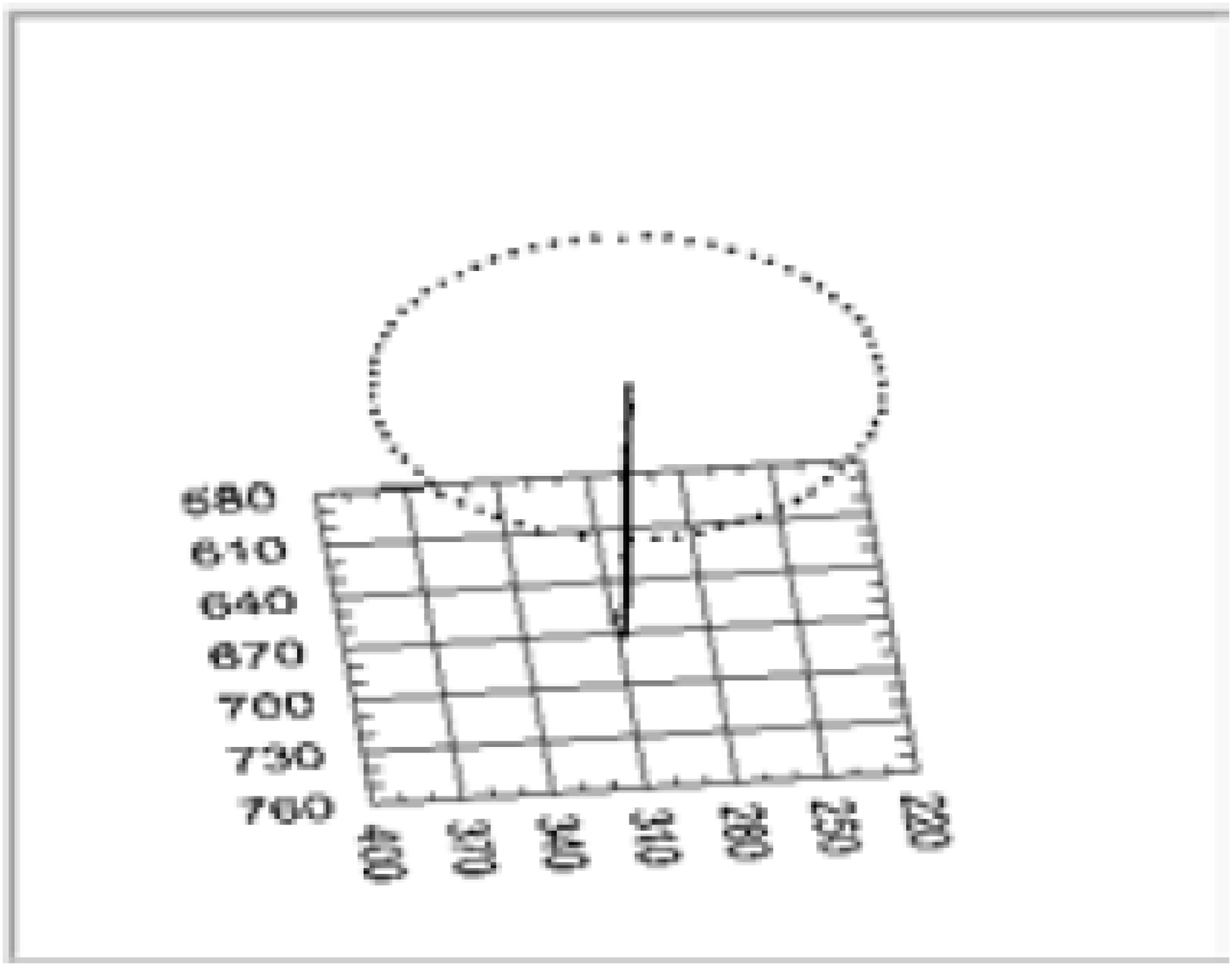}\\
  \caption{Axis and Edges of Section $\sharp130$ of the straw tube$\sharp4$ in Tomo reference.
  The Straw axis is at 9 degree respect to the tomograph vertical axis (the figure axis is not
  in scale respect the horizontal ones)}
  \end{center}
\end{figure}

\begin{figure}
\begin{center}
  \includegraphics[width=8cm]{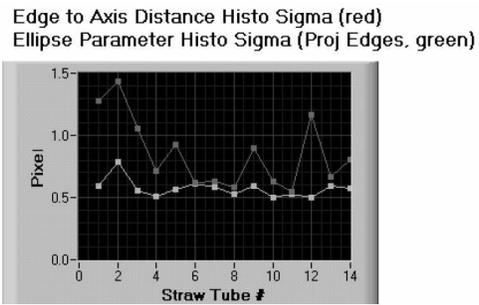}\\
  \caption{Standard deviation of the histogram of the Edge radius (red) and sigma of the gaussian
  fit to the ellipse parameters of the projected edges (green) for 14 straw tubes.}
  \end{center}
\end{figure}

\begin{figure}
\begin{center}
  \includegraphics[width=5cm]{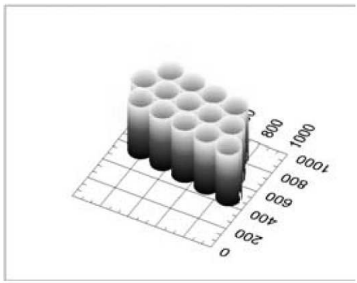}\\
  \caption{3-D reconstruction in Tomo reference of the 14-straw module.}
  \end{center}
\end{figure}
\newpage
\section{  \textsc{Conclusions}}
We have developed a new technique to visualize lw-mass surface of
cylindrical shapes widely used in HEP detectors, such as straw
tubes. The technique uses x-ray computed tomography, implemented
with an original optical recognition, pattern recognition and
analysis code, Labview-based. Preliminary results show how the
precision of our technique in determining deviation from circular
shapes are better than $30\mu$m.
\newline
%

\newpage

\end{document}